\documentclass[aip,jcp,superscriptaddress,citeautoscript,longbibliography,floatfix,twocolumn,10pt]{revtex4-2}
\usepackage{graphicx}
\usepackage{amsmath}
\usepackage{amssymb}
\usepackage{url}

\renewcommand{\epsilon}{\varepsilon}
    
\graphicspath{{Figures/}}
\usepackage{bm}
\usepackage{url}
\usepackage{breakurl}
\usepackage[breaklinks]{hyperref}

\usepackage{color}

\newcommand{\tn}[1]{\textnormal{#1}}  
\newcommand{\bluee}[1]{{#1}}
\newcommand{\blue}[1]{{#1}}

\begin{document}

\title{\Large Collapse and expansion kinetics of a single polyelectrolyte chain with hydrodynamic interactions} 

\author{Jiaxing Yuan}
\affiliation{Research Center for Advanced Science and Technology, The University of Tokyo, 4-6-1 Komaba, Meguro-ku, Tokyo 153-8904, Japan}

\author{Tine Curk}
\email{tcurk@jhu.edu}
\affiliation{Department of Materials Science and Engineering, Johns Hopkins University, Baltimore, Maryland 21218, USA}

\begin{abstract}
We investigate the collapse and expansion dynamics of a linear polyelectrolyte (PE) with hydrodynamic interactions. 
Using dissipative particle dynamics with a bead-spring PE model, long-range electrostatics and explicit ions we examine how the timescales of collapse~$t_\text{col}$ and expansion~$t_\text{exp}$ depend on the chain length~$N$, and obtain scaling relationships $t_\text{col}\sim N^\alpha$ and $t_\text{exp}\sim N^\beta$. 
For neutral polymers, we derive values of $\alpha=0.94\pm0.01$ and $\beta=1.97\pm0.10$.
Interestingly, the introduction of electrostatic interaction markedly shifts $\alpha$ to $\alpha\approx1.4\pm0.1$ for salt concentrations within $c=10^{-4}~\tn{M}$ to $10^{-2}~\tn{M}$. A reduction in ion-to-monomer size ratio noticeably reduces~$\alpha$.  
On the other hand, the expansion scaling remains approximately constant, $\beta \approx 2$, regardless of salt concentration or ion size considered. We find $\beta > \alpha$ for all conditions considered, implying that expansion is always slower than collapse in the limit of long polymers.  \blue{This asymmetry is explained by distinct kinetic pathways of collapse and expansion processes.}
\end{abstract}

\maketitle

Polyelectrolytes (PE) are charged polymers that constitute a class of materials with profound significance in
biological systems (such as proteins, DNA, and RNA) and diverse
industrial
applications~\cite{rubinstein2012polyelectrolytes,murugappan2022physics}. 
A substantial body of theoretical and computational research has been
dedicated to unraveling the structural attributes of these
polymers~\cite{dobrynin2005theory,rubinstein2012polyelectrolytes} 
and gaining insights into
their expansion-collapse
transition~\cite{lee2001dynamics,lee2004collapse,ghosh2021kinetics}. In
particular, the non-equilibrium kinetics
governing the PE expansion-collapse transition holds vital
implications for elucidating the folding dynamics of
biomolecules~\cite{bloomfield91, russell2002rapid,yuan2024Impact} and for designing
responsive smart materials~\cite{tagliazucchi15,li2022soft}.

The non-equilibrium phase ordering kinetics of soft materials can be 
significantly influenced by hydrodynamic interactions (HI) induced by solvent flow~\cite{Manghi2006,gompper2009multi,Marchetti2013hydrodynamics,Yuan2022Impact,yuan2023mechanical}. \blue{For a
neutral polymer in a solvent a collapse can be induced by a change in the solvent quality, in which case HI were shown to accelerate the
collapse kinetics  compared to Brownian dynamics (BD)
simulations~\cite{chang2001solvent,abrams2002collapse,Kikuchi2005,Kamata2009}. }
Starting from the fully expanded chain Kikuchi \emph{et al.} provided an analytical prediction asserting that the collapse timescale~$t_\text{col}$ follows a power-law relationship, $t_\text{col}\sim N^{4/3}$, with $N$ representing the chain length~\cite{Kikuchi2005}.
Guo \emph{et al.} employed dissipative particle dynamics (DPD) simulations and reported a different scaling behavior, $t_\text{col}\sim N^{0.98\pm0.09}$, if the initial condition is an equilibrated polymer.

However, it is not clear how the presence of electrostatic interactions impacts the scaling relationship of PE collapse. 
The full hydrodynamic simulation of the expansion--collapse kinetics of charged PE is complex due to the necessity of properly accounting for long-range and many-body electrostatic interactions as well as HI. Such investigations are relatively rare with only one recent computational study~\cite{yuan2023HI} delving into PE collapse kinetics using fluid particle dynamics method~\cite{Tanaka2000}, which revealed that HI significantly speeds up PE collapse but did not explore how HI changes the scaling behavior. 
Moreover, the dependence of expansion timescale $t_\text{exp}$, the reverse process of collapse, on the chain length~$N$ has to our knowledge not been investigated.

In this work, we conduct DPD
simulations of a coarse-grained (CG) bead-spring PE~\cite{stevens93,
  stevens95} to investigate how the collapse time~$t_\text{col}$ and
expansion time~$t_\text{exp}$ scales with the chain
length~$N$. Our CG model comprises an anionic PE,
monovalent counterions, and additional monovalent salt at concentration $c$ contained in a
cubic three-dimensional periodic box of size $L$. The PE is
represented as a bead-spring chain~\cite{stevens93, stevens95}
composed of $N$ monomers, each carrying a charge of $-e$. The PE
monomers and ions are treated as spherical particles with diameter
$\sigma$ and $\sigma_\mathrm{s}$, respectively. We primarily focus on
the case of $\sigma=\sigma_\mathrm{s}$, but we also explore the impact of a smaller ion size ($\sigma_\mathrm{s}=0.5\sigma$). The
particles interact through a standard 12-6 Lennard-Jones (LJ) potential
characterized by an energy coupling constant $\varepsilon$ and a cutoff
distance $r_\mathrm{cut}$. We choose $r_\mathrm{cut}=3\sigma$ to
represent the short-range hydrophobic attraction between monomers, whereas
$r_\mathrm{cut}=2^{1/6}(\sigma+\sigma_\mathrm{s})/2$ and
$r_\mathrm{cut}=2^{1/6}\sigma_\mathrm{s}$ are set for purely repulsive monomer--ion and
ion--ion LJ interaction.
The electrostatic coupling is controlled by the Bjerrum length~
$l_{\mathrm{B}}$ given by $l_{\mathrm{B}}= e^2/(4\pi
k_\mathrm{B}T\varepsilon_\mathrm{sol})$ where
$\varepsilon_{\mathrm{sol}}$ represents the permittivity of the
solvent, $k_{\mathrm{B}}$ is Boltzmann's constant, and $T$ is the
absolute temperature. We set
$\sigma=l_{\mathrm{B}}=0.72~\mathrm{nm}$ to represent a typical monomer size and electrostatic coupling in an aqueous electrolyte at room temperature~\cite{stevens95,stevens98}.
Neighboring monomers along a chain are connected through
harmonic potential $U_{\mathrm{bond}}(r_{ij})=(K/2)({r_{ij}}-R_0)^2$
with spring constant $K=400k_{\mathrm{B}}T/{\sigma}^2$ and bond
length $R_0=2^{1/6}\sigma$, where $r_{ij}$ represents the center-to-center distance between monomer~$i$ and $j$.
Electrostatic interactions are calculated using the
particle-particle-particle-mesh (PPPM) algorithm with a relative force
accuracy of $10^{-3}$~\cite{darden93,deserno1998mesh}.
Each simulation contains a single polymer chain in a box size of $L=120\sigma$ for $N\le100$
and $L=180\sigma$ for $N=200$, which is sufficiently large to avoid polymer--polymer interaction through periodic images \blue{(see Table~S1 and Fig.~S1 in Supplemental Information (SI) for details).}

To account for hydrodynamic interactions, we employ the computational method based on
dissipative particle dynamics (DPD)~\cite{grootwarren} that couples the polymers to the DPD solvent (DPDS)~\cite{curk2024dissipative}. 
The monomers and ions are immersed in a DPD
solvent characterized by parameters typical for an aqueous
solution: particle density $\rho=3 r_{\mathrm{c}}^{-3}$, cutoff distance $r_{\mathrm{c}}=\lambda=0.646~\mathrm{nm}$, friction parameter
$\gamma=4.5 k_{\mathrm{B}} T \tau / r_{\mathrm{c}}^2$, with $\tau=\lambda\sqrt{m/(k_\tn{B}T})$ the standard molecular dynamics simulation time unit, $m$ the mass of the particles, and interaction prefactor $a_{ij}=78 k_{\mathrm{B}} T$; see
Ref.~\onlinecite{curk2024dissipative} for more details). The coupling between the solute particles (monomers and ions) and the DPD solvent
is achieved by friction parameter $\gamma_{\mathrm{s}}=5 \gamma$ and cutoff distance $r_{\mathrm{s}}=r_{\mathrm{c}}$.
This setting captures the hydrodynamic interactions and yields the typical diffusion constant of ions and monomers  $D \approx 0.07\lambda^2/\tau  \approx 1.1 \mathrm{~nm}^2 / \mathrm{ns}$ \blue{($\tau\approx0.027\mathrm{ns}$)}.
The system is evolved using the
velocity-Verlet integrator with a time step of $\Delta t=0.005 \tau$.
When plotting the results, we scale the time in terms of the Brownian time for a free particle
$\tau_\mathrm{BD}=\sigma^2/(24D)\approx0.74\tau$.

We first perform equilibrium simulations to prepare initial equilibrated configurations of the polymer.
When investigating the collapse process, we initially employ purely
repulsive LJ interactions between the polymer segments ($r_\tn{cut}=2^{1/6}\sigma$ and $\varepsilon=3k_{\mathrm{B}} T$) while
incorporating full electrostatic interactions. To initiate the
collapse, we introduce an attractive LJ interaction between the
monomers by setting $r_\tn{cut}=3\sigma$ and keeping $\varepsilon=3k_{\mathrm{B}} T$. This attractive interaction is activated instantaneously, leading
to a sudden quench that drives the polymer collapse. Conversely, when
studying expansion, we follow the opposite protocol. The equilibrium structures are generated using attractive LJ interactions ($r_\tn{cut}=3\sigma$) with $\varepsilon=3k_{\mathrm{B}} T$. Subsequently, we change the monomer--monomer LJ interaction to a purely repulsive ($r_\tn{cut}=2^{1/6}\sigma$) to investigate the expansion kinetics.

\begin{figure*}[h!]
  \centering \includegraphics[width=12cm]{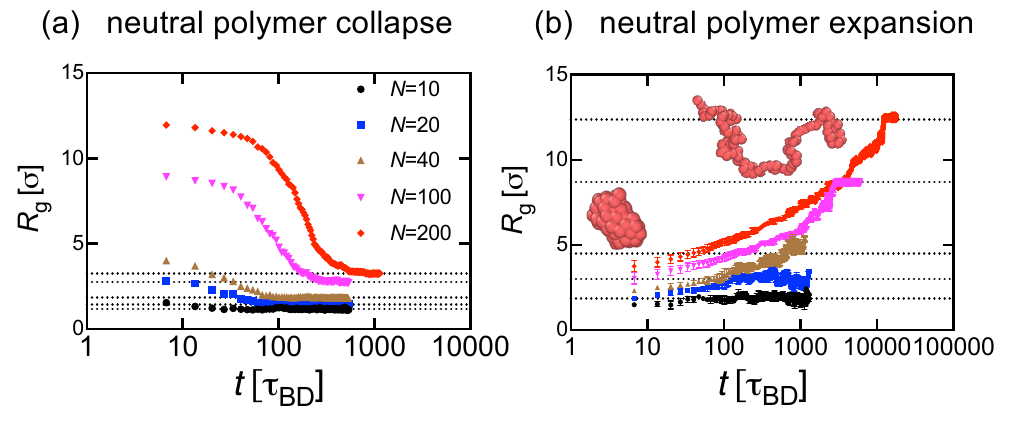}
  \caption{The temporal change of $R_{\mathrm{g}}$ as a function of
    elapsed time~$t$ for a single neutral polymer of different chain lengths~$N$ undergoing (a) collapse and (b) expansion. The dashed lines show the equilibrium collapsed (a) and expanded (b) $R_\tn{g}$.
    The insets in (b) show the equilibrium collapsed and expanded polymer configurations for $N=100$. Error bars denote standard errors obtained from 10 independent simulations. Error bars in (a) are smaller than the symbol size.}
    \label{fig:Figure1}
\end{figure*}

\begin{figure*}[h]
  \centering \includegraphics[width=12cm]{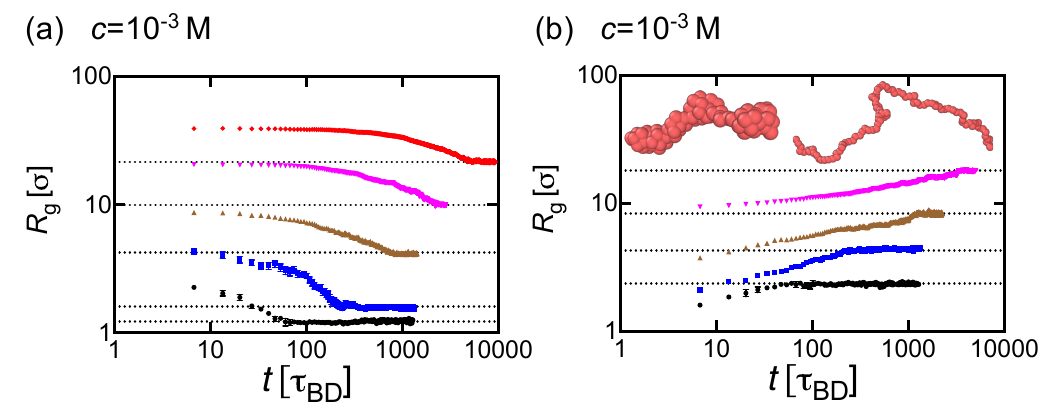}
  \caption{The temporal change of $R_{\mathrm{g}}$ for a single PE of various chain lengths~$N$
    undergoing collapse (a) and expansion (b) \blue{at salt concentration $c=10^{-3}\,\tn{M}$}.
    The insets in (b) show the representative equilibrated collapsed and expanded PE configurations for $N=100$.
    Salt ions are not shown for clarity.
    The monomer size and ion size are identical ($\sigma_\mathrm{s}=\sigma$). Error bars show standard errors, for most points they are smaller than the symbol size.
    }
    \label{fig:Figure2}
\end{figure*}

Using this setup, we examine the collapse and expansion dynamics of both neutral polymer and charged PE \blue{at different salt concentrations $c=10^{-4}~\tn{M}$ (Debye length $l_D\approx30.4\mathrm{nm}$), $10^{-3}~\tn{M}$ ($l_D\approx9.61\mathrm{nm}$), and $10^{-2}~\tn{M}$ ($l_D\approx3.04\mathrm{nm}$).}
The temporal change of $R_{\mathrm{g}}$ as a function of elapsed
time~$t$ for the collapse and expansion of a single neutral polymer and a charged PE \blue{at salt concentration $c=10^{-3}\,\tn{M}$} are shown in Fig.~\ref{fig:Figure1} and Fig.~\ref{fig:Figure2}, respectively. 
\blue{The raw data of PE collapse and expansion at different salt concentrations $c=10^{-4}\,\tn{M}$ and $c=10^{-2}\,\tn{M}$ are provided in Fig.~S2 in SI.}
Comparing PE with a neutral polymer, we observe that the electrostatic repulsion between monomers significantly slows the collapse kinetics, while
expansion is accelerated. This
effect becomes particularly important when the salt concentration is
low since electrostatic repulsion is not effectively screened.
Moreover, the equilibrium configurations depend sensitively on the electrostatic repulsion and salt concentrations~$c$: neutral polymer forms a globule, but the PE forms an elongated rod or a bead-spring configuration with larger $R_\tn{g}$ at lower $c$, in agreement with previous studies~\cite{limbach2003single,hsiao06}.
This change in equilibrium $R_\tn{g}$ significantly affects the collapse and expansion timescales in addition to direct electrostatic interactions.

\begin{figure*}[h]
  \centering \includegraphics[width=12cm]{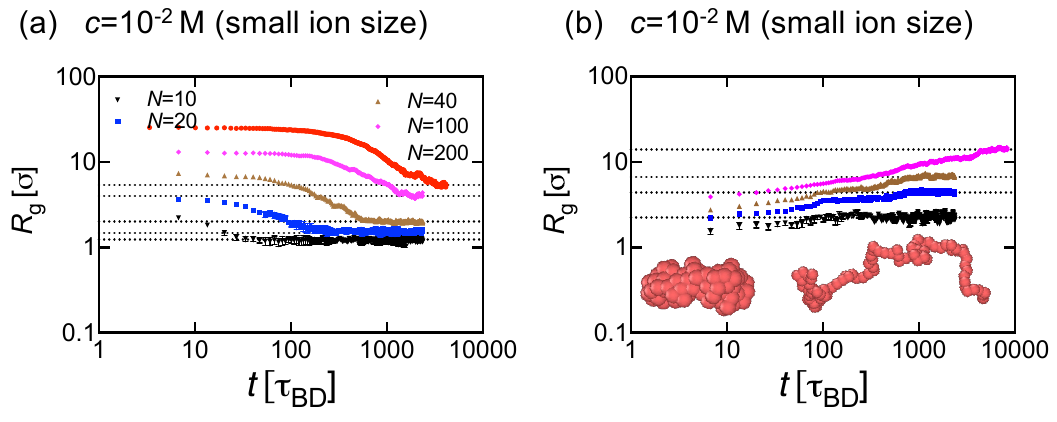}
  \caption{The temporal change of $R_{\mathrm{g}}$ for a single PE
    undergoing (a) collapse and (b) expansion at $c=10^{-2}M$ and smaller ion size ($\sigma_\mathrm{s}=0.5\sigma$).
    The insets in (b) show the collapsed and expanded PE for $N=100$.}
    \label{fig:Figure3}
\end{figure*}

\begin{figure*}[h]
  \centering \includegraphics[width=18cm]{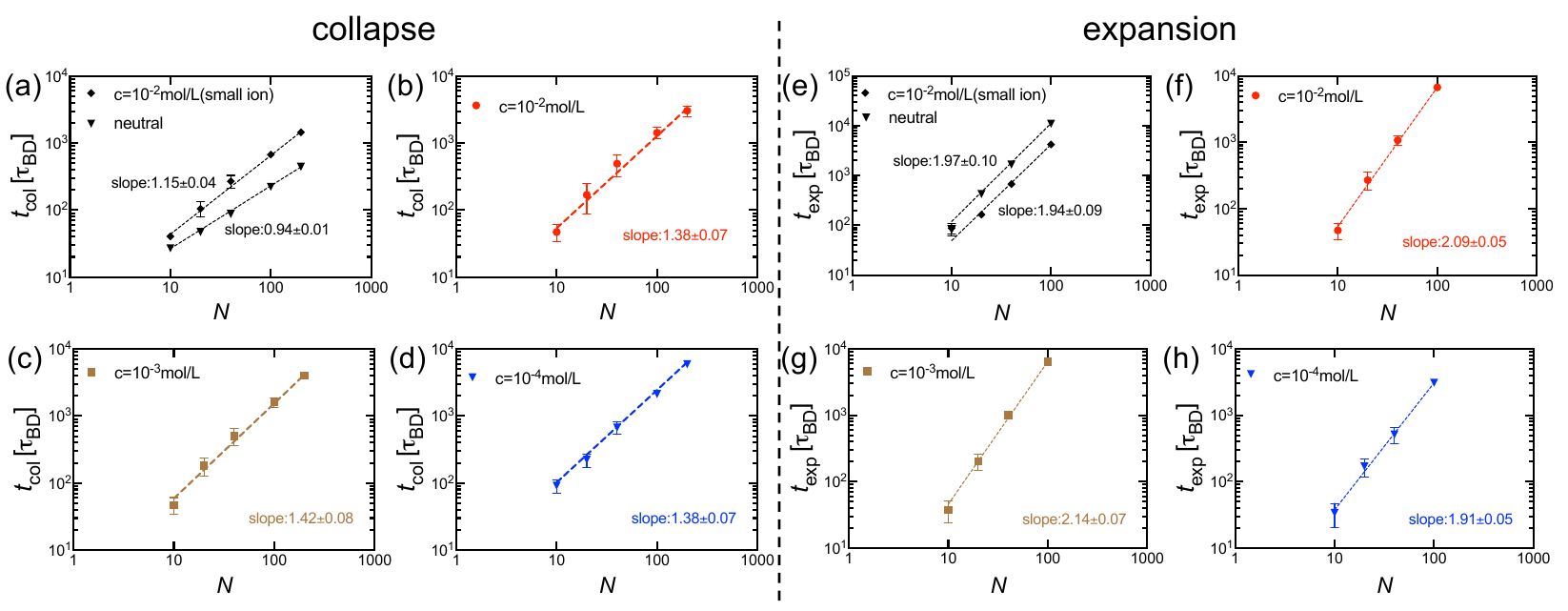}
  \caption{Collapse and expansion timescales of charged polymer and neutral polymer calculated by applying Eq.~\eqref{eq:Rgt} to data in Figs.~\ref{fig:Figure1}--\ref{fig:Figure3}.
The exponents for collapse~$\alpha$ and expansion~$\beta$ are obtained by fitting the data to 
 $t_\text{col}\propto N^{\alpha}$ and  $t_\text{exp}\propto N^{\beta}$ respectively.
   }
      \label{fig:Figure4}
\end{figure*}

For these simulations (Fig.~\ref{fig:Figure2}) we employed the standard choice with salt ions of the same size as the monomers, $\sigma_\tn{s}=\sigma$. However, ion size can impact ion binding, as it determines the closest distance 
between ions and monomers.
In Fig.~\ref{fig:Figure3}, we explore the influence of ion
size. It is evident that a smaller ion size ($\sigma_\tn{s}=0.5\sigma$) yields more
compact structures in the collapsed PE due to stronger
electrostatic attraction between ions and monomers, consistent with
previous simulations of PE~\cite{hsiao06,yuan19dielectric,yuan2020structure}. 
Compared to PE collapse where ions and monomers have identical sizes \blue{(Fig.~S1(b))}, smaller ion size substantially accelerates the collapse  
since electrostatic repulsion between monomers is screened more effectively. Moreover, smaller ions can more easily incorporate into the globular polymer thus yielding compact equilibrium globular structures that are similar to neutral polymer configurations  (compare configurations in Fig.~\ref{fig:Figure3}b with \blue{Fig.~S1(d)}and Fig.~\ref{fig:Figure1}b)

Based on the time evolution of $R_{\mathrm{g}}$ in Fig.~\ref{fig:Figure1}--Fig.~\ref{fig:Figure3}, 
we calculate how the collapse and expansion
timescale depend on the chain
length~$N$. 
The collapse and expansion timescale $t_\text{col}$ and $t_\text{exp}$
are defined by the time \bluee{$t^*$} required for the change in the radius of
gyration $R_{\mathrm{g}}(t)$ to reach a fraction $f_{\mathrm{c}}$ of the
maximum change~\cite{Kikuchi2005},
\begin{equation}
R_{\mathrm{g}}(t^*)=R_{\mathrm{g},
  \mathrm{init}}-f_{\mathrm{c}}\left(R_{\mathrm{g},
  \mathrm{init}}-R_{\mathrm{g}, \mathrm{final}}\right),
  \label{eq:Rgt}
\end{equation}
where $R_{\mathrm{g}, \mathrm{init}}$ and $R_{\mathrm{g},
  \mathrm{final}}$ are, respectively, the initial and the final equilibrated values
of $R_{\mathrm{g}}$.
In the present work, we set $f_{\mathrm{c}}=0.9$ and fit the obtained timescales to a power-law scaling ($t_\text{col}\sim N^\alpha$ and
$t_\text{exp}\sim N^\beta$) using the nonlinear Levenberg-Marquardt algorithm~\cite{marquardt1963algorithm} \blue{that takes into account the error bars when fitting the data. We employ the implementation of this algorithm in Gnuplot}~\cite{Gnuplot}.
\blue{For the collapse process, we conduct 15 independent simulations. Conversely, for the expansion process, owing to the absence of driving force and larger variance, we conduct a range of 25 to 35 independent runs. This enables us to acquire high-resolution data concerning $R_\mathrm{g}(t)$ and its associated timescale. Note the choice of $f_{\mathrm{c}}=0.9$ ensures the extracted timescales are not affected by the slow local arrangements of the collapsed globule in the late stage.}~\cite{Kikuchi2005}

The scaling behaviors of collapse and expansion timescales are summarized in
Fig.~\ref{fig:Figure4}. For a neutral polymer we obtain $\alpha=0.94\pm0.01$ (Fig.~\ref{fig:Figure4}(a)) and 
$\beta=1.97\pm0.10$ (Fig.~\ref{fig:Figure4}(e)). This agrees with previous reported
simulation results~\cite{guo2011coil} of neutral polymer collapse starting from an initially equilibrated conformation, $t_\text{col}\sim N^{0.98\pm0.09}$. 

Notably, we find the collapse scaling for PE is $\alpha\approx1.4\pm0.1$ (Fig.~\ref{fig:Figure4}b--d) for different salt concentrations, indicating that the incorporation of electrostatic interactions not only shifts
the absolute values of timescales but also markedly changes the scaling behavior compared to the neutral polymer case. We speculate that $\alpha\approx1.4\pm0.1$ scaling may be a result of the initial state resembling a fully expanded polymer with $R_{\mathrm{g}}\sim N$ due to electrostatic repulsion between segments. Consequently, $\alpha\approx1.4\pm0.1$ is close to the theoretical prediction of $\alpha=4/3$ for neutral polymer collapse starting from a fully expanded state~\cite{Kikuchi2005}. We speculate that scaling could change to lower values of $\alpha$ for very long polymers $N>200$ and high salt concentration $c>10^{-2}$~M. Conversely, the expansion scaling does not change appreciably with the introduction of electrostatics and we find  $\beta \approx 2$ (Fig.~\ref{fig:Figure4}f--h) for the charged PE under different salt concentrations.

In the case of smaller ion-to-monomer size ratio, $\sigma_{\tn s}/\sigma=0.5$, we obtain $\alpha=1.15\pm0.04$ at $c=10^{-2}$~M (Fig.~\ref{fig:Figure4}a), which is closer to the corresponding values for neutral polymers, while the expansion scaling remains unaffected within statistical accuracy of our results, $\beta=1.94\pm0.09$ (Fig.~\ref{fig:Figure4}e). We attribute the reduced $\alpha$ to enhanced screening and the ability of small ions to be incorporated into the globule, enabling a charge-neutral globule configuration. In contrast, with larger ion-to-monomer size ratio ($\sigma_{\tn s}/\sigma=1$), the ions cannot incorporate into a collapsed globule, leading to more expanded, pearl-necklace-like equilibrium structures. These findings suggests that the collapse scaling is sensitive to the short-range ion--polymer interaction and whether ions can incorporate in, or are excluded from, the collapsed polymer globule. \bluee{We anticipate that the scaling exponent $\alpha$ varies continuously with the size ratio $\sigma_{\tn s}/\sigma$ and that by further decreasing the size ratio, the strengthened screening would result in $\alpha$ closer to that of a neutral polymer ($\alpha\approx 1$).}

\begin{figure*}[h]
  \centering \includegraphics[width=11cm]{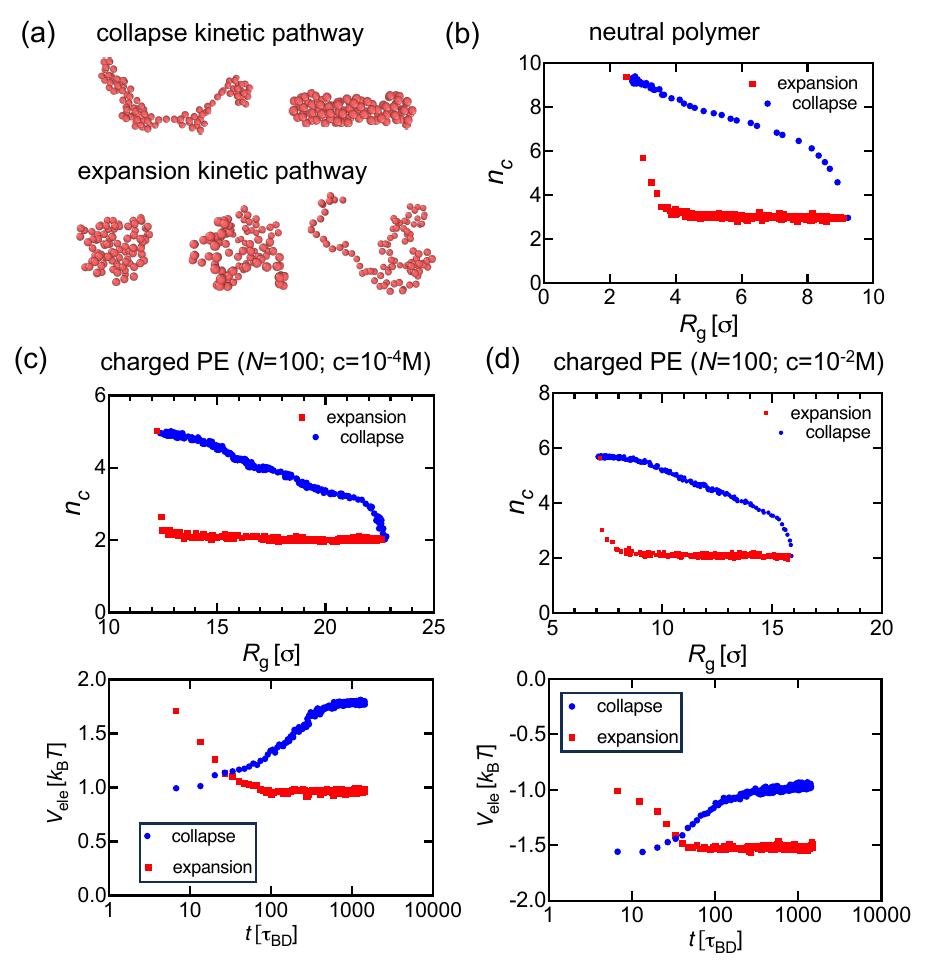}
  \caption{\blue{Characterization of kinetic pathways in collapse and expansion processes for a polymer and PE with $N=100$. (a) Snapshots from simulations depicting polymer collapse and expansion kinetic pathways for a neutral polymer.
  (b) Relationship between the radius of gyration~$R_\mathrm{g}$ and the number of contacts per monomer~$n_\mathrm{c}$ during collapse and expansion of a neutral polymer.
  (c)--(d) Top panels: The same analysis as in panel (b) conducted for the collapse and expansion of a charged PE at salt concentrations of $c=10^{-4}~\text{M}$ and $c=10^{-2}~\text{M}$.
  Bottom panels: The temporal change of the average electrostatic energy per charged particle (i.e., PE and salt ions), denoted as $V_\mathrm{ele}$, for the collapse and expansion of a PE at salt concentrations of $c=10^{-4}~\text{M}$ (bottom left) and $c=10^{-2}~\text{M}$ (bottom right). }
   }
      \label{fig:Figure5}
\end{figure*}

Interestingly, we find $\beta > \alpha$ for all conditions investigated, implying that expansion is always slower than collapse for long polymers. \blue{This intriguing observation can be understood by considering the kinetic pathways for collapse and expansion (Fig.~\ref{fig:Figure5}a). The pathway is quantified by a $R_\text{g}$--$n_c$ parametric plot where $n_\tn{c}$ is the number of contacts per monomer defined as the average number of neighbors within a distance $1.5\sigma$ (Fig.~\ref{fig:Figure5}b). Notably, we find that a closed circle emerges confirming that collapse and expansion follow distinct kinetic pathways. The collapse behavior resembles a shrinking process along the backbone where small pearls form whose hydrodynamic drag is proportional to the pearl diameter ($N^{1/3}$) and in addition HI induce directional flow accelerating the merging of local pearls~\cite{Kikuchi2005,yuan2023HI}. Conversely, during expansion the coil expands isotropically where the total drag scales as the sum over monomers ($N$) and there is no directional flow along the backbone, resulting in a longer timescale compared to the collapse.}

Furthermore, while collapse scaling is substantially affected by electrostatics and ion--polymer interaction, the expansion scaling remains $\beta\approx2$ regardless of polymer charge, salt concentration and ion size (Fig.~\ref{fig:Figure4}e--h). This suggests that the expansion scaling is governed by the diffusive expansion of the polymer and electrostatic repulsion does not appear to noticeably alter this scaling, even at very low salt concentrations.
\blue{We analyze the $R_\text{g}$--$n_c$ plot for the collapse and expansion of PE under salt concentrations of $c=10^{-4}$~M (Fig.~\ref{fig:Figure5}c, top) and $c=10^{-2}$~M (Fig.~\ref{fig:Figure5}d, top). For expansion, we observe a two-step process: initially, the coil experiences slight swelling, during which the number of contacts per particle~$n_\tn{c}$ is significantly reduced. Subsequently, diffusion-driven expansion occurs, constituting the rate-limiting stage, which is only weakly influenced by electrostatics due to the relatively small monomer contacts $n_\tn{c}$. Conversely, $n_\tn{c}$ continues to rise during the entire collapse process, which underscores the critical role of electrostatics on the collapse.
\bluee{The structural analysis of radial density distribution for polymer and charged PE during collapse and expansion collectively support the above scenario (Fig.~S3 in SI).}
These findings also align with the evolution of electrostatic energy~$V_\mathrm{ele}$ (Fig.~\ref{fig:Figure5}c--d, bottom) where $V_\mathrm{ele}$
reaches saturation much earlier in the expansion compared to the collapse. \bluee{Note $V_\mathrm{ele}$ is defined for the entire system encompassing monomers, counterions, and salt ions and its value depends on the relative number of ions to monomers in the system. $V_\mathrm{ele}$ is positive in the low-salt regime, which is dominated by monomer--monomer repulsion, but it becomes negative in the high salt case where electrostatics is dominated by ion--ion attractive interactions. The meaningful aspect lies in the temporal change of $V_\mathrm{ele}$ during the collapse/expansion process, which is related to the conformational change of the PE.}
The different kinetic pathways of collapse and expansion can explain why the collapse scaling $\alpha$ is sensitive to electrostatics, while the expansion scaling, $\beta\approx2$, is not.}

\blue{Following previous works on polymer collapse~\cite{chang2001solvent,abrams2002collapse,Kikuchi2005,Kamata2009,majumder2017kinetics,yuan2023HI}, we immediately quench the system to initiate collapse or expansion. For a typical real PE with chain length $N\approx10^4$, based on the predicted scaling relationships $t_\text{col}\sim N^\alpha$ and $t_\text{exp}\sim N^\beta$, we infer that the timescales for collapse and expansion under the salt concentration of $c=10^{-2}$M are $t_\text{col}\approx13\mu s$ and $t_\text{exp}\approx1350 \mu s$.
Thus, our work should be relevant to experimental conditions where the rapid quenching into a new regime can be achieved within microseconds.}

To summarize,  we systematically examined the collapse and expansion
timescales of neutral polymers and charged polyelectrolytes with long-range electrostatics, explicit salt ions and hydrodynamic interactions. Our findings
illustrate that the inclusion of electrostatic interactions
alters the scaling behaviors $t_\text{col}\sim N^\alpha$ of collapse, while it does not noticeably affect the expansion scaling
$t_\text{exp}\sim N^\beta$. Specifically, in neutral polymer systems, we obtain
$\alpha=0.94\pm0.01$,
consistent with the previous prediction~\cite{guo2011coil} and $\beta=1.97\pm0.10$.
The inclusion of electrostatic interactions significantly shifts the value of collapse scaling to $\alpha\approx1.4$.
Interestingly, changes in salt concentration ranging from $10^{-4}~\tn{M}$ to $10^{-2}~\tn{M}$ do not have a substantial impact on the scaling relationships of both collapse and expansion timescales. However, reducing ion-to-monomer size ratio changes the scaling to $\alpha\approx1.15$ at $c=10^{-2}~\tn{M}$, which we mainly attribute to the ability of small ions to incorporate into the collapsed polymer globule. This highlights that non-electrostatic short-range ion--polymer interaction can substantially influence the collapse scaling.  The expansion scaling, however, remains constant with $\beta\approx2$ regardless of the charge, salt concentration or ion size. We find that $\beta > \alpha$ for all conditions considered implying that expansion is always slower than collapse for long polymers. 
\blue{We elucidate these observations by contrasting the distinct kinetic pathways of collapse and expansion (Fig.~\ref{fig:Figure5}), highlighting that these two processes are not simply opposites of each other.}
Our study provides fundamental
insights into the collapse and expansion kinetics of linear polyelectrolytes, useful for rationalizing the dynamics of polyelectrolytes in solution
and understanding the folding and unfolding process of charged biopolymers.

\section{Supplementary Material}
Supplemental material provides the raw $R_\tn{g}(t)$ data used to calculate the collapse and expansion timescales, system-size comparison and simulation details. 

\begin{acknowledgments}
  This work was supported by the startup funds provided by the Whiting School of Engineering at JHU and performed using the Advanced Research Computing at Hopkins (rockﬁsh.jhu.edu), which is supported by the National Science Foundation (NSF) grant number OAC 1920103.
\end{acknowledgments}

%

\end{document}